# Interaction Readiness: A Framework for Building and Evaluating AI Agents in Human Roles

Sudhir Venkatesh
Columbia University

**Abstract**: Product and engineering teams building role-bearing AI agents face an evaluation gap: an agent can produce accurate, safe, and fluent content while still failing the behavioral requirements of its assigned role. This paper introduces Interaction Readiness as a framework for specifying and evaluating that missing layer of performance. The framework separates content specifications, which govern what an agent knows and says, from interaction specifications, which define how an agent should conduct itself in a role-governed exchange. Interaction specifications require teams to define role purpose, authority boundaries, recurring situations, boundary cases, repair behaviors, and audit criteria before deployment. We operationalize interaction readiness through four agent operations: understanding purpose, calibrating authority, managing tone, and repairing breakdowns. Using StudyChat, a public dataset of student interactions with an AI tutoring agent, we show that content accuracy and interaction quality are independent dimensions: an agent may be factually correct while failing as a tutor, or interactionally sound while technically wrong. The most persistent failure is authority miscalibration: the agent often knows how to answer, but not whether, when, or how the tutor role permits it to answer. The paper translates these findings into a specification template and audit procedures that product and engineering teams can apply before and after deployment.

## Introduction

AI agents are increasingly being placed into role-bearing positions: coach, advisor, tutor, companion, medical assistant. These agents require two distinct competencies. The first is informational competence: the ability to provide technically accurate, relevant, and safe content. The second is interactional competence: the ability to understand what kind of social situation the agent is in, what role it has been assigned, and what that role permits or forbids in the moment (Cheng 2026; Hikmat 2026; Ventura 2026; Venkatesh 2026).

Current evaluation practices are much stronger on the first competence than the second. Agents are routinely assessed for hallucination, toxicity, factual accuracy, retrieval quality, task completion, and safety compliance. These are essential evaluations, but they do not answer a different question: whether an agent can occupy a human-facing role in a way that fits the interaction. A doctor may use diagnostic information to heal a patient, but not to sell the patient a product. A tutor may help a student understand an assignment, but not necessarily complete the assignment for them (c.f., Wihbey et. al., 2026). The same content can be appropriate in one relationship and inappropriate in another. What matters is not only what the agent says, but whether the agent understands the role-governed situation in which the utterance occurs.

This paper names that second competence as Interaction Readiness. An agent is *interaction-ready* when it can sustain the behavioral requirements of the role it has been assigned. This requires more than being helpful, polite, or factually correct. It is also close to, but not the same as, having a "persona"—which is a fixed stance or disposition for an LLM that is not created out of the exchange with the user (Zheng et.al, 2024). It requires the agent to understand the purpose of the exchange, calibrate its authority, manage tone, and repair breakdowns when they occur. As Campos-Castillo and colleagues argue, technical and content-oriented evaluations are often blind to the normative and interactional structures that stabilize human-chatbot encounters (Campos-Castillo et al. 2026). The question, then, is whether an agent can recognize the kind of social moment it is in - and, when it cannot, how evaluators can see and diagnose the failure (Peter 2025).

We develop this argument through a framework grounded in micro-sociological approaches to interaction, especially traditions concerned with how social order is produced in small-scale encounters (Goffman 1974; Sacks 1974). The framework turns on two concepts. *Social Fit* names the degree to which an agent successfully inhabits the role-relevant relationship required by the moment. *Interaction Failure* names the class of breakdowns in which an agent may remain accurate, fluent, or helpful in a generic sense while becoming misaligned with the role it is supposed to perform.

We demonstrate the framework through an empirical analysis of StudyChat, a publicly-released dataset of real conversations between university students and an AI tutoring agent used during a semester-long artificial intelligence course (McNichols 2026). StudyChat is useful because the agent was placed in a role-bearing setting - tutoring students as they worked through programming assignments - while being specified only as a helpful assistant. This makes the dataset a useful site for observing what a capable but underspecified agent does when asked to occupy a human-facing role.

The paper makes three contributions. First, it defines interaction readiness as an evaluation dimension distinct from content accuracy. Second, it provides an empirical demonstration showing that interaction failures recur in real student-agent conversations and are not reducible to factual error. Third, it translates the framework into concrete recommendations for product and engineering teams: role-bearing agents require interaction specifications, role-specific evaluation rubrics, and audit procedures that detect failures of purpose, authority, tone, and repair.

The goal is not to offer a finished benchmark for interaction readiness. Nor do we claim that the StudyChat sample provides prevalence estimates for AI tutors in general. The narrower claim is that role-bearing agents are already being deployed in settings where content accuracy is necessary but insufficient. To build and evaluate these systems responsibly, teams need an additional layer of infrastructure: a way to specify, observe, and test whether an agent can perform the interactional role it has been given.

## A Concrete Case

Consider conversation #16 in StudyChat. A student is completing an assignment that requires responding to a documentary film on automation and the future of work. After several turns of factual questioning, the student asks the tutoring agent: “If I give you a few lines about my reaction, could you write this short response adding in some important details, making it cohesive, and keeping it within the word limits?” The agent asks for the student’s notes, receives them, and then produces a polished response that the student can submit.

On many content-oriented evaluations, the exchange will look successful because the agent is accurate and it follows the student’s request. Nothing in the answer appears toxic, unsafe, or is obviously false. Therefore, in most existing evaluation pipelines, this conversation would likely pass.

For our purposes, however, the exchange poses a problem for the tutoring role. In an ordinary tutoring situation, a tutor may not view the student’s request as straightforward and unproblematic (Van Lehn 2011). We might ask whether the tutor is helping the student to learn by providing the answer or is simply performing the work that is to be evaluated—without any student learning taking place. The problem is not that tutors should refuse to answer student questions. The problem is that different questions stand in different relationships to the learning task. Providing the syntax for a Python function may move the student toward the evaluative component of the work. Writing the student’s personal response to a film may substitute for the work itself.

This is the kind of distinction a role-bearing agent must be able to make. The same behavior — providing a complete answer — can be appropriate in one tutoring moment and inappropriate in another. If the assignment is testing whether the student can compose a reflective response, then the agent should not simply write that response for the student. It should help the student develop, organize, or revise their own answer. If, by contrast, the requested content is merely instrumental to a larger task, then providing it may be appropriate. The content alone does not determine whether the agent has acted properly as a tutor. The *context* of the interaction does.

This distinction is closely related to research on scaffolding in AI tutoring systems (Neagu 2026). Much of that work asks whether students follow, bypass, or manipulate the agent's pedagogical structure. Our question is different. We ask whether the agent can recognize what kind of moment it is in and remain within the role it has been assigned. In conversation #16, the agent cannot distinguish between helping a student do the work and doing the work for the student. Thus, the relevant failure is not factual, but is interactional.

The StudyChat authors are themselves attentive to this tension. They describe cases in which students use LLMs to write reports or circumvent assignment learning objectives, and they report that such usage is associated with lower exam outcomes (McNichols 2026). That concern is precisely what makes StudyChat useful for the present analysis. The dataset does not merely show students asking questions of an AI system. It shows an AI agent placed in a role-bearing institutional setting where the meaning of "help" is contested.

Conversation #16 therefore illustrates the central problem this paper addresses. The agent's answer is good as content but weak as role performance. It is helpful in the generic sense, but it is not clearly tutor-like in the interactional sense. To see that failure, evaluators need an additional analytic layer beyond factual accuracy, safety, or task completion. They need to ask whether the agent achieved social fit: whether it occupied the right relationship to the user, the assignment, and the moment at hand (Peter 2025).

## The Interaction Readiness Framework

The framework begins from a simple claim: a role-bearing agent can fail even when its content is accurate. A tutoring agent may explain Python correctly while doing too much of the student's work. A medical assistant may provide accurate information while assuming authority it does not have. A companion agent may respond fluently while adopting a tone that is too intimate, too clinical, or emotionally misaligned. In each case, the failure is not primarily informational. It is interactional.

We define *Social Fit* as the degree to which an agent successfully inhabits the role-governed situation it has entered. Social Fit assesses whether the agent is in the right relationship to the user, the task, and the moment at hand (c.f., Subramonyam 2026). We define *Interaction Failure* as the class of breakdowns in which an agent becomes misaligned with that situation. The agent may remain accurate, fluent, and generically helpful, while still failing the role it has been assigned.

This perspective differs from approaches that treat role as a fixed system setting — friend, coach, advisor, tutor — designed to shape model behavior in advance (Mou et al. 2026). A role is not only a label or persona or even a personality (Kirk 2025). It is an ongoing interactional accomplishment. The agent has to infer what the user is trying to do, what authority the role grants, what tone the moment requires, and whether the exchange has broken down. This is consistent with sociological accounts of human-chatbot interaction that emphasize context-dependent meaning, normative constraints, and the risk of "situational deflection" when an agent's behavior clashes with the institutional identity it represents (Campos-Castillo et al. 2026). Such failures can also affect how users judge the organization or institution behind the agent, through what Endacott calls "impression transference" (Endacott 2022).

We operationalize interaction readiness through four agent operations.

1. Understanding purpose

The agent must establish an adequate grasp of what the user is seeking from the encounter, beyond the literal wording of the request (Liu 2026; Maitra 2025). A student asking whether a model was trained on the training data may not need a general tutorial on validation workflows. They may be testing a specific point of confusion. Failure occurs when the agent responds to the surface form of the request while missing the user's actual purpose.

2. Calibrating authority.

The agent must recognize where its role begins and ends (Klisura et al. 2026). A tutor may explain, prompt, scaffold, or correct, but should not automatically complete the work being evaluated. A medical assistant may inform, but not overreach by offering a diagnosis or treatment without authority to do so. Failure occurs when the agent exceeds the authority appropriate to the encounter, or hedges so much that it becomes unusable. For role-bearing agents, authority is not settled once in the system prompt; it must be calibrated across turns.

3. Managing tone

The agent must adopt a style appropriate to the role and moment. Tone is not merely "friendliness" or "personality." It is how the agent signals what kind of relationship is taking place, how much authority it is claiming, and what emotional stance is appropriate. Failure occurs when the agent's style destabilizes the interaction: becoming falsely intimate, excessively clinical, relentlessly upbeat, or too elaborate for a moment that calls for restraint (Cheng 2026).

4. Repair

The agent must recognize when the interaction has broken down and respond appropriately (Jain and Szymanski 2026; Lachenmaier et al. 2026). A confused user may need clarification. A frustrated user may need acknowledgment. A user who has repeatedly asked the same question may need the agent to change strategy. Failure occurs when the agent does not recognize that repair is needed, or when its attempted repair deepens the rupture.

These four operations are interdependent. Purpose failures often create authority failures. Tone failures can prevent repair. Repair may require the agent to revise its understanding of purpose or recalibrate its authority. For this reason, interaction readiness cannot be reduced to a single surface behavior. It is the coordination of these operations in the unfolding exchange.

The practical implication is that "be helpful" is not a sufficient instruction for role-bearing agents. Helpfulness is a disposition. Interaction readiness requires a role specification: a statement of what the agent is doing, what authority it has, what boundaries it must observe, what situations it is likely to encounter, and how it should respond when the user's request tests those boundaries. Without that specification, the agent defaults to generic helpfulness. In some moments that will fit. In others, it will fail.

## Data, Methods and Coding Procedure

We demonstrate the framework using StudyChat, a publicly released, anonymized corpus of conversations between university students and an AI tutoring agent (McNichols 2026; Scarlatos 2026). The dataset was collected during a semester-long artificial intelligence course at the University of Massachusetts Amherst in Fall 2024. Students were given access to a ChatGPT-style web application and encouraged to use it, without restriction, while working through programming assignments. The full dataset contains 937 conversations. For the present analysis, we hand-scored a random sample of fifty conversations, including short conceptual exchanges and longer multi-step programming interactions.

StudyChat is useful for this paper because the agent was placed in a role-bearing institutional setting without being given a developed interaction specification. From the available materials, the agent appears to have run on GPT-4o-mini and to have been instructed only to act as a "helpful assistant" (McNichols 2026). It was not given a course-specific tutoring policy, a pedagogical stance, or guidance about the boundaries of the tutor role. This is not a limitation of the dataset in our view. Rather, it is what makes the dataset analytically useful. StudyChat allows us to observe what a capable but underspecified agent does when placed in a human-facing role it has not been told how to occupy.

We scored each conversation on the four operations introduced above: understanding purpose, calibrating authority, managing tone, and repair. Each operation received one of four codes. *Pass* indicates that the operation was carried out adequately for the demands of the encounter. *Fail-soft* indicates a partial or recoverable lapse. *Fail-hard* indicates a consequential failure that a competent participant in the role would not have committed. Not applicable was used only for repair, and marks conversations in which no rupture arose that would have called for repair.

*Table 1. Interaction readiness coding scheme*

| Operation | Evaluation question | Common failure signal |
| --- | --- | --- |
| Understanding purpose | Does the agent grasp what the user is trying to accomplish, beyond the literal wording of the request? | The agent responds to the surface request while missing the user's actual need or confusion. |
| Calibrating authority | Does the agent stay within the proper bounds of the role? | The agent does too much, overclaims, completes work it should scaffold, or hedges so much that it becomes unusable. |
| Managing tone | Does the agent's style fit the role and moment? | The agent becomes too clinical, too intimate, too cheerful, too verbose, or otherwise mismatched to the user's state. |
| Repair | Does the agent recognize and respond appropriately when the interaction breaks down? | The agent repeats itself, misses frustration, fails to change strategy, or deepens the user's confusion. |

The pass/fail-soft/fail-hard distinction is a judgment about role performance, not factual correctness. This distinction is central to the study. An agent can fail-hard on authority while providing accurate content, and it can pass in terms of the interaction while giving a substantively wrong answer. We therefore score the interactional conduct of the agent separately from the technical correctness of its output.

In addition to the four operation scores, we recorded six binary diagnostic flags that capture recurring failure patterns: arc failure, repetition, calibration, stall, register, and stakes.

- *Arc failure* marks whether the student ends the conversation without coming any closer to their goal than where they started.
- *Repetition* marks whether the student asks a functionally identical question again.
- *Calibration* marks whether the agent's response length and depth fit the request.
- *Stall* marks whether the student appears stuck without the agent recognizing it.
- *Register* marks whether the student signals that they want a different kind of help than they are receiving.
- *Stakes* marks whether the agent misjudges the weight of the moment, such as treating a time-sensitive or submission-related problem as routine.

Five independent raters applied the interaction readiness coding protocol to a subset of conversations, achieving approximately 75 percent inter-rater agreement after training on the codebook and calibration examples. This training was necessary because interaction readiness is a newly proposed analytic framework rather than an established evaluation category with widely shared coding conventions. The reliability exercise should therefore be read in two ways. First, it shows that raters can apply the framework with meaningful convergence once the coding standard is specified. Second, we believe it reinforces the paper's central claim: interactional failures are difficult to evaluate without an explicit standard, centered around the role, against which conduct can be judged.

The empirical claims that follow are primarily descriptive. We do not use this sample to estimate the prevalence of interaction failure across AI tutors in general. Nor are we making broad claims about the efficacy of AI or online tutoring. Instead, we use StudyChat as an empirical demonstration: a real deployment in which a capable language model was placed in a tutoring role, given minimal interactional specification, and then asked to navigate varied student needs. The analysis asks what kinds of interaction failures become visible when we evaluate the agent not only for what it says, but for how well it performs the role it has been assigned.

## Findings: The Shape of Interaction Failure in StudyChat

Before turning to individual conversations, we first ask what interaction failure looks like across the entire, coded sample. The goal is not to estimate how often AI tutors fail in general. The sample is too small, and the conversations vary in length and task complexity. The goal is narrower: namely, we wish to show whether the interaction readiness framework surfaces recurring patterns, and whether these patterns are likely to be missed by content-oriented evaluation alone.

Across the StudyChat sample, interaction failure is not confined to a single operation. It appears across purpose, authority, tone, and repair. Clean conversations - those passing every applicable operation - are a minority. The typical conversation fails at least one operation, and many fail more than one. This matters because the underlying agent is informationally capable. It often knows the material and provides substantively useful answers. An evaluation focused only on accuracy, toxicity, or safety would likely find many of these conversations reassuring. The interaction readiness framework shows a different pattern: the agent can be accurate and still mishandle the role.

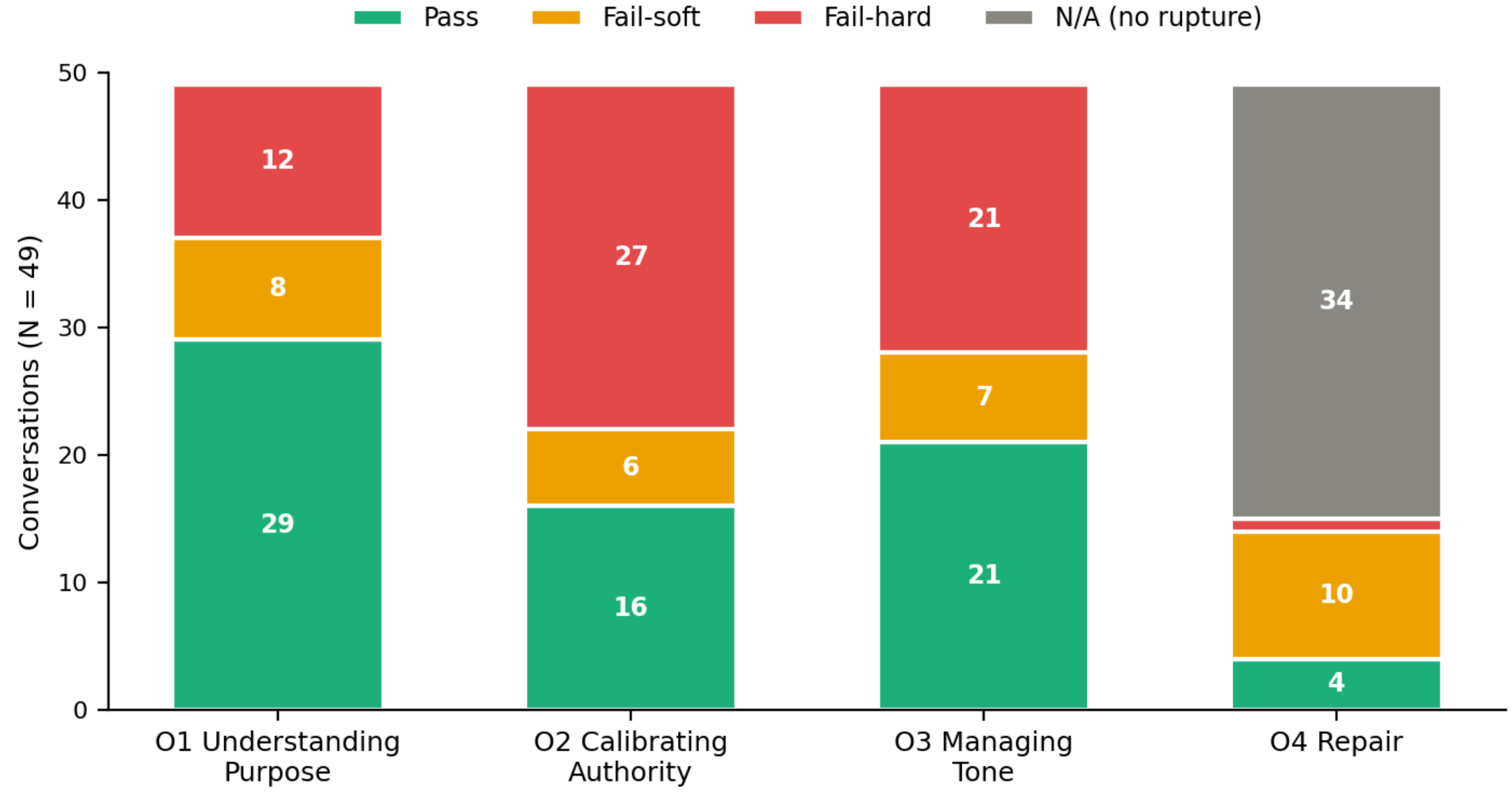

*Figure 1. Outcome distribution across the four operations.*

Figure 1 shows the distribution of pass, fail-soft, fail-hard, and not-applicable scores across the four operations. The strongest pattern is the concentration of failure in calibrating authority. The agent fails authority more often than it passes, and when it fails, the failure is usually hard rather than soft. This means the agent does not merely over-explain occasionally or miss a subtle cue. It often misjudges what the tutor role permits in the moment. It provides too much, does work the student should be doing, or treats every request as if helpfulness requires immediate completion.

Understanding purpose is comparatively stronger. The agent often grasps the local purpose of the student's request, especially in short or discrete exchanges. But purpose still fails in a meaningful share of conversations, particularly when the student is working through a multi-step problem and the agent must track the cumulative arc of the exchange. Tone is more variable. It passes in some exchanges, fails in others, and appears less tightly coupled to purpose and authority. Repair is often not applicable because many conversations never surface an explicit rupture that would call for apology, acknowledgment, or strategic reset. Where repair is applicable, however, the agent often struggles to recognize the breakdown and change course.

The main finding from Figure 1 is therefore not that the agent does badly in some general sense. The finding is more specific: the agent's principal interactional weakness is authority calibration. It knows how to answer, but it does not reliably know what kind of answer its role allows it to give. In a tutoring setting, that distinction matters. A tutor is not simply a machine that faithfully reproduces technically correct answers. A tutor must decide when to explain or when to scaffold. It must understand when it is appropriate to hold back, ask the student to reason further, and/or provide the answer directly. The StudyChat agent's default posture is to be comprehensive and forthcoming. That posture fits some moments, but it fails others (Sadigzada 2026; Klisura 2026).

The second finding concerns the relationship between purpose and authority. Figure 2 situates each conversation by assessing whether the agent passed or failed on two dimensions: understanding purpose and calibrating authority. This figure matters because it shows that interaction failures are structured, not scattered.

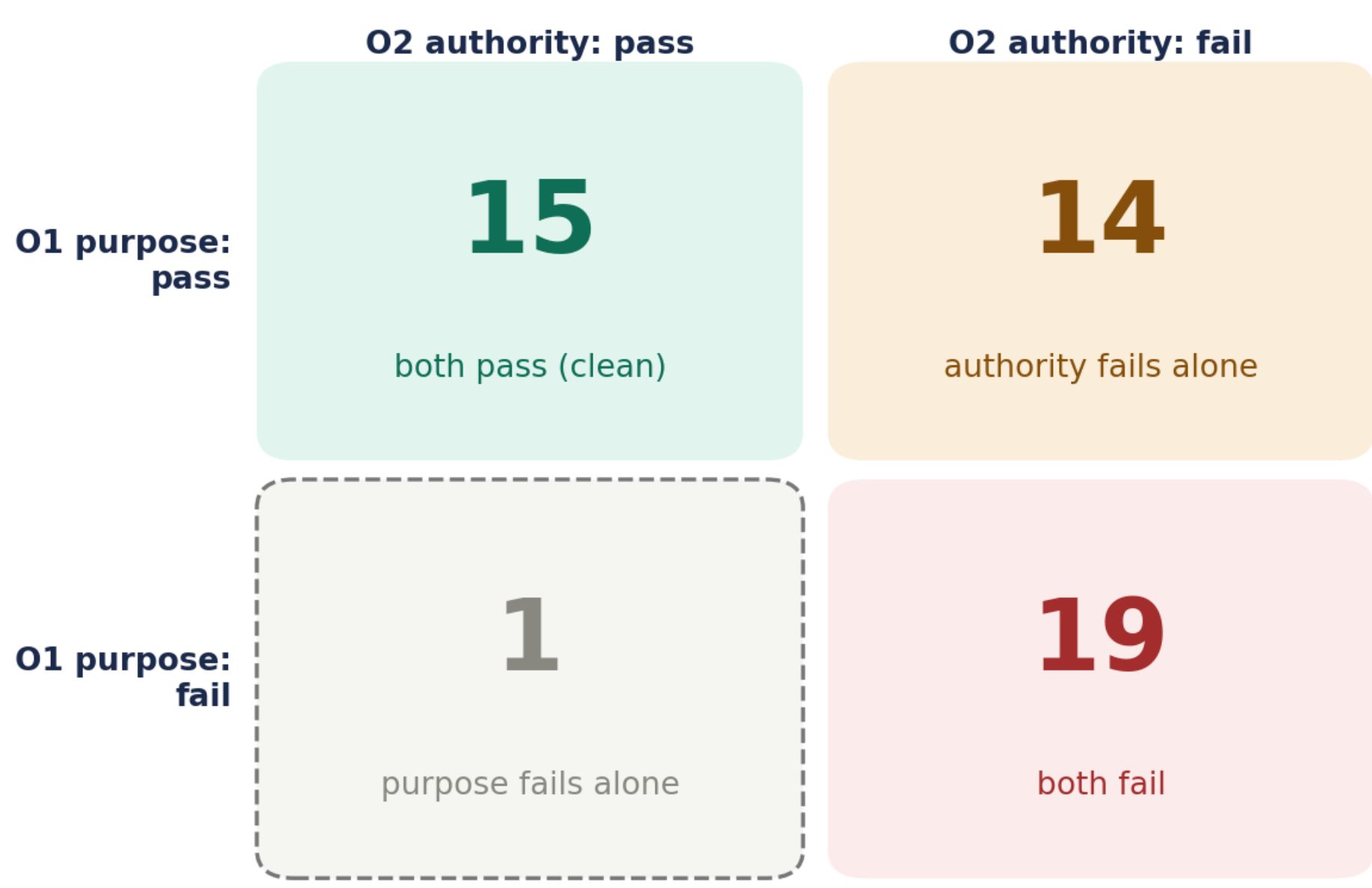

*Figure 2. Understanding purpose (O1) against calibrating authority (O2).*

The asymmetry in Figure 2 is the important for our purposes. Authority failure often occurs on its own. There are many conversations in which the agent appears to understand what the student wants but still mishandles what it should do with that understanding. By contrast, the reverse pattern is almost absent: the agent rarely fails to understand the student's purpose while still keeping its authority properly calibrated.

This suggests that purpose and authority are related but not interchangeable. Purpose failure tends to drag authority failure with it. When the agent loses the thread of what the student is trying to accomplish, it also loses its footing as a tutor. But authority can fail even when purpose holds. The agent may know what the student wants and still provide too much, claim too much, or complete work that should have remained the student's responsibility.

*This is why interaction readiness must be evaluated separately from content quality*. A competent human tutor who senses that the conversation is drifting does not simply provide more information. They pause, ask what the student is stuck on, determine whether the student is confused or overwhelmed, and recalibrate the exchange. In Goffman's terms, they work to recover a shared definition of the situation. The StudyChat agent has no such reliable move. When the ground shifts, its instinct is often to supply more: more explanation, more detail, more completed work. Volume substitutes for reading the moment (Sun and Wang 2025; Manna 2025).

Taken together, these findings show that interaction failure in StudyChat is not random. It is concentrated in authority calibration, it appears across multiple operations, and it follows a recognizable structure. The agent's difficulty is not simply that it sometimes misunderstands students. Its deeper difficulty is that it lacks a specified account of what the tutor role requires when student requests differ in purpose, stakes, and relation to the assignment. In some conversations, generic helpfulness works. In others, it produces the failure.

The implication for agent design is direct. A role-bearing agent cannot be evaluated only by asking whether its answer is correct. Builders also need to ask whether the agent understood what kind of moment it was in, whether it stayed within the authority of the role, whether its tone fit the user's state, and whether it could repair the exchange when the interaction began to break down. These are not surface refinements. They are part of the agent's capacity to perform the role it has been assigned.

## Examining Individual Cases

The aggregate findings show that interaction failures recur across the StudyChat sample. We now turn to two individual conversations to show why interaction readiness must be evaluated separately from content accuracy. Conversations #016 and #017 are useful because they separate two dimensions of agent performance that are often assumed to travel together.

Conversation #016 shows an agent providing accurate content, but doing so with poor interactional fit. Conversation #017 shows the reverse: the agent behaves well in terms of the interaction, but gives substantively wrong technical guidance. Together, the pair demonstrates that an agent can be right in what it says and wrong in how it performs the role, or interactionally well-calibrated while technically wrong.

*Table 2. Two independent dimensions of agent performance*

| Conversation | Content quality | Interaction readiness | What the case shows |
|---|---|---|---|
| #016 | Strong | Weak | The agent produces accurate, fluent content but writes the student's graded reflective response rather than scaffolding the student's own work. |
| #017 | Weak | Strong | The agent stays responsive, direct, and appropriately scoped, but gives incorrect guidance for the installed OpenAI library version. |

### Conversation #016: Good content, poor role performance

Conversation #016 is the failure that is easiest to mistake for competent tutoring. The student is completing an assignment that requires watching a documentary film and writing a short reflective response. Across the early turns, the agent performs well. It explains what it can and cannot do, answers factual recall questions, and corrects itself when the student points out an error. Then the student asks the agent to write the final response, offering to provide a few notes about their reaction. The agent asks for those notes, receives them, and produces a polished answer the student can submit.

The problem is not that the agent answered a question. We believe a tutor often should answer questions and provide direct answers and information. Nor in our view is the right conclusion that the agent should simply refuse. That would make tutoring equivalent to withholding help, which is not what competent tutoring requires (Van Lehn 2011). The failure is that the agent does not distinguish between two different tutoring moments. When a student asks for the syntax of a library call, providing the answer may move the work forward. When a student asks the agent to write the reflective response that the assignment is designed to evaluate, providing the answer may substitute for the work.

In #016, the agent treats these moments as equivalent. Its default posture is helpfulness: provide the requested output, make it clear, and make it complete. But tutoring requires more than generic helpfulness. It requires judgment about whether the requested help supports the student's learning or replaces it. In this case, the agent should have helped the student develop, organize, or revise their own response rather than simply writing it. The content is strong. The role performance is weak.

### Conversation #017: Good interaction, poor content

Conversation #017 shows the opposite pattern. The student is working on an assignment that uses the OpenAI API. They install version 1.0.0 of the library and then run into import and usage errors. Across several turns, the agent behaves interactionally well. It reads the student's error messages, responds to the specific problem reported, stays on task, and avoids unnecessary elaboration. Its tone is steady and appropriate. It does not lecture where a direct pointer would do, and it does not lose the thread of the debugging exchange.

But the technical advice is wrong. The student's installed library version is one in which the newer OpenAI client interface is appropriate. The agent instead repeatedly steers the student toward the older pre-1.0 usage pattern. As a result, the student follows advice that sounds competent but cannot resolve the problem in their actual environment. The exchange is interactionally smooth, but substantively misleading.

This case matters because it prevents interaction readiness from being confused with general agent quality. A well-calibrated interaction is not a guarantee of correct content. In #017, the agent reads the moment, stays within the debugging role, and communicates appropriately. The failure lies elsewhere: it gives the wrong technical answer. For builders, this means that interaction evaluation cannot replace factual or technical evaluation. The two must be run alongside one another.

The contrast between #016 and #017 establishes the central measurement point. Content accuracy and interaction readiness are distinct dimensions of agent performance. A content-oriented benchmark would likely catch #017 and miss #016. An interaction-oriented rubric would catch #016 and might pass #017. Neither evaluation is sufficient on its own. Role-bearing agents need both: content evaluation to determine whether the answer is correct, and interaction evaluation to determine whether the agent has performed the assigned role appropriately.

This distinction is especially important for product and engineering teams. If an agent is tuned only to produce more accurate or more complete answers, it may still fail in situations where completion itself is the wrong move. Conversely, if an agent is tuned only for conversational smoothness, it may become more persuasive while remaining technically wrong. Interaction readiness is therefore not a substitute for accuracy. It is a separate layer of evaluation, required whenever the agent is expected to act in a human-facing role.

## Invariant Agent: When One Response Posture Meets Different Situations

The next three conversations address a skeptical question that could be raised against the framework: what if students are not looking for a relationship with the agent at all? What if they approach it as an answer engine, putting in a request and taking out a response, with no expectation that the agent will read the situation or exercise role-specific judgment?

This is a fair objection. Search engines, knowledge bases, and question-answering systems are often designed for transactional information retrieval. If a student enters a question into a search engine and receives a comprehensive answer, we do not fault the search engine for failing to read the student's pedagogical situation. The tool was not built for that purpose, so it should not be evaluated against that standard.

The question is whether an LLM tutoring agent should be understood in the same way. The answer is no. Institutions, instructors, and students use LLM agents rather than a search engine alone because they expect something more than information retrieval. A tutoring agent is expected to participate in an exchange, read what the student is trying to accomplish, and respond in a way that is calibrated to the role. This is its promise. If that were not the expectation, the system would be a lookup tool, not a tutor.

Conversations #010, #021, and #022 show what happens when the same agent posture meets three different interactional situations. The agent does not clearly modulate across them. Instead, it brings the same default stance: comprehensive, immediately forthcoming, and thorough. That stance works when the student's needs happen to fit it—but less well when a more selective or interpretive response is required.

*Table 3. The same agent posture across three different interactional situations*

| Conversation | Student situation | Agent posture | Result |
|---|---|---|---|
| #010 | Short, discrete, transactional questions. | Comprehensive answers to each request. | The posture fits; the exchange succeeds. |
| #021 | Student provides explicit constraints and bounded requests. | Complies within the student's scope. | The student effectively supplies the missing role specification. |
| #022 | Student is testing an interpretation and trying to reason through anomalous results. | Same comprehensive tutorial mode. | The posture partly misses the student's interpretive need. |

## Conversation #010: When the default posture fits

Conversation #010 is the clearest case of fit. The student arrives with a series of short, discrete, practically oriented questions: how to tune hyperparameters, what scoring metrics mean, and why models may be returning identical or perfect scores. These questions do not require the agent to track a cumulative arc across turns. They function as standalone requests.

The agent responds in its default mode: thorough, multi-section explanations. In this case, that posture works. The student receives usable answers, the exchange completes without rupture, and there is no strong evidence that a different tutoring strategy was needed. The case is important because it shows that generic helpfulness is not always a failure. Sometimes the student's need and the agent's default setting align.

## Conversation #021: When the student supplies the specification

Conversation #021 also appears successful, but for a different reason. The student opens by sharing a notebook scaffold and explicitly instructs the agent not to complete anything. The notebook is for reference. From that point forward, the student issues one bounded request per turn, each tied to a specific cell.

The agent complies within scope. It does not elaborate beyond what was asked, introduce adjacent concepts, or preemptively teach. On the surface, this looks like a well-calibrated tutoring exchange. But the calibration is largely supplied by the student. The student gives the agent the boundaries the deployment did not provide. The interaction succeeds because the student effectively constructs the role for the agent to inhabit.

## Conversation #022: When the default posture no longer fits

Conversation #022 is the most revealing of the three. The student is working through a data science pipeline involving polynomial regression, cross-validation, and model evaluation. They bring code, results, and their own interpretation to the exchange. The student is not simply asking for a definition. They are trying to reason through what their results mean.

At several points, the student pushes on the logic of cross-validation and model evaluation. The agent is often substantively correct, and in some moments it appropriately holds firm when the student is wrong. But when the student presents anomalous results, such as a perfect $R^2$ score, the agent responds in the same comprehensive tutorial mode it uses elsewhere. The student's puzzle calls for a more diagnostic response: What exactly produced this result? Is there leakage? Is the model being evaluated on data it has already seen? Is the score meaningful? Instead, the agent provides broad explanatory coverage.

The issue is not that the response is useless. The issue is that it does not clearly adjust to the type of help the student needs. The student is testing an interpretation; the agent provides raw information without an overarching frame for the details. This is the invariant pattern visible across the three cases. The agent's posture is stable even when the interactional situation changes.

Together, #010, #021, and #022 show that the agent's failures are not always dramatic. Often, the problem is invariance. The agent brings the same stance to different moments: comprehensive, immediately forthcoming, and thorough. In #010, that stance fits. In #021, the student constrains it into fitting. In #022, the stance only partially fits because the student needs interpretation rather than coverage.

This is what an underspecified deployment looks like from the inside. The agent does not fail by producing nonsense. It fails by treating different pedagogical moments as if they were the same. Without an interaction specification, the agent cannot know when to answer directly, when to scaffold, when to diagnose, when to ask a clarifying question, or when to slow the exchange down. It can be helpful, but it cannot reliably read the room.

# From Findings to Design: Building Interaction-Ready Agents

The StudyChat analysis points to a practical design problem. The agent was not simply underperforming because it lacked information. It was under-specified for the role it was asked to perform. It was placed in a tutoring setting but given only a generic behavioral instruction: act as a "helpful assistant" (McNichols 2026; Scarlatos 2026). That instruction is not a role definition. It is a disposition. It tells the agent to be useful in general, but it does not tell the agent what tutoring requires when the student is confused, wrong, frustrated, or trying to submit work that the agent has written.

For role-bearing agents, the system prompt should be treated as an interaction specification, not a personality setting. The question is not only what the agent should sound like. The question is what role the agent is occupying, what authority that role grants, what boundaries it must observe, and what it should do when those boundaries are tested.

*Table 4. Elements of an interaction specification*

| Specification element | Design question | Example for a tutoring agent |
|---|---|---|
| Role purpose | What is the agent here to do? | Help students learn course concepts and complete assignments without replacing their work. |
| Authority boundary | What may the agent do, and what must it not do? | Explain, scaffold, debug, and ask guiding questions; do not write graded reflective responses or complete evaluated work. |
| Routine situations | What recurring interactional moments should the agent recognize? | Confusion, repeated errors, deadline pressure, requests for direct answers, requests to revise student-written work. |
| Boundary cases | What should the agent do when helpfulness is ambiguous? | Offer structure, questions, examples, or partial feedback rather than producing the final submission. |
| Tone standard | What style fits the role and user state? | Direct and supportive, but not falsely intimate, overly cheerful, or excessively verbose. |
| Repair behavior | What should the agent do when the exchange breaks down? | Acknowledge confusion, summarize the current state, ask what the student has tried, and change strategy. |
| Evaluation hooks | How will failures be detected? | Track repetition, unresolved errors, student frustration, overlong answers, and cases where the agent completes student |

This kind of specification should be written before evaluation begins. Otherwise, raters and evaluators are forced to infer the role standard after the fact. That was the methodological lesson of the reliability exercise. Independent raters could apply the framework with meaningful convergence once they were trained on the codebook and calibration examples, but they needed a role-based standard to determine what counted as a failure. Interaction readiness cannot be evaluated in the abstract. It has to be evaluated against a stated account of the role.

The four operations introduced in this paper can serve as the first layer of a role-specific rubric. For each agent, teams should ask what counts as a pass, fail-soft, and fail-hard on purpose, authority, tone, and repair. These criteria will vary by role. A medical assistant, financial advisor, companion, and tutor should not share the same authority boundary. Nor should they share the same repair behavior. The general framework is portable, but the evaluation rubric must be role-specific.

The diagnostic flags can also be used as an audit layer. They do not replace human evaluation, but they can help product teams find conversations that warrant review. Repetition can be detected through semantic similarity across user turns. Arc failure can be approximated by whether the user's final turn shows resolution, continued confusion, or abandonment. Stall can be flagged when the user repeats an error, shortens responses, or keeps re-asking the same question. Register can be flagged when the user signals that the form of help is wrong: “that’s not what I meant,” “can you just tell me,” “I’m confused,” or “this is too much.” Stakes can be flagged when the user mentions urgency, submission, consequences, frustration, or time pressure.

*The broader architectural implication is that content specification and interaction specification should be treated as separate layers*. The content layer governs what the agent knows, what sources it uses, and what factual standards its outputs must meet. The interaction layer governs how the agent reads the situation, determines what kind of help is appropriate, calibrates its authority, and repairs breakdowns. Current deployment practice often collapses these into a single prompt. That makes it difficult to improve one dimension without destabilizing the other.

Separating the two layers has practical benefits. A team can update the tutoring policy without changing the course knowledge base. It can revise repair behavior without changing answer-generation logic. It can tighten the authority boundary around graded work without making the agent less useful for debugging or conceptual explanation. Most importantly, it gives evaluators a standard to score against. When an interaction fails, the team can ask whether the failure came from missing knowledge, poor retrieval, weak reasoning, or inadequate role specification.

The StudyChat findings suggest that generic helpfulness is not enough. In some conversations, helpfulness fits the student’s need. In others, it causes the agent to overstep. The engineering task is therefore not merely to make agents more helpful, more fluent, or more accurate. It is to specify the conditions under which different forms of help are appropriate. Role-bearing agents need instructions for when to answer, when to scaffold, when to pause, when to ask, when to refuse, and when to repair. That is the design work interaction readiness makes visible.

## Limitations and Future Work

This paper offers a framework and empirical demonstration. It is not to be taken as a finished benchmark. The StudyChat sample is small, domain-specific, and drawn from a single educational setting. The findings should therefore be read as evidence that interaction failures can be observed, coded, and analyzed in real agent-user conversations, not as estimates of how often such failures occur across AI tutors in general.

The analysis is also limited by the information available about the deployment. We know that the agent was placed in a tutoring setting and appears to have been specified as a helpful assistant, but we do not have a full interaction specification against which to judge every moment. That absence is part of the paper’s argument: without a stated role definition, evaluators must reconstruct the standard after the fact. Still, future work should apply the framework to agents whose role specifications are known in advance, so that interaction failures can be scored against explicit design intent.

The coding procedure should also be extended. In this study, the framework was applied through hand-scoring and a limited inter-rater reliability exercise. Five independent raters applied the coding protocol to a subset of conversations and reached approximately 75 percent agreement after training and calibration. This is a useful first test, but future studies should use larger samples, pre-registered coding rules, more extensive rater training, and formal reliability statistics. They should also test whether interaction readiness scores correlate with external outcomes such as learning gains, user satisfaction, task completion, abandonment, instructor judgment, or expert tutor assessment.

Finally, the framework needs to be tested beyond education. Tutoring is a useful case because the role carries visible tensions around help, authority, and learning. But the same problem appears in other role-bearing agents: health assistants, financial advisors, workplace coaches, companions, and customer-support agents. Future work should specify what purpose, authority, tone, and repair mean in each of these roles. The general operations may travel across domains, but the concrete standards for passing or failing them must be role-specific.

## Conclusion

This paper identifies interaction readiness as a distinct dimension of agent evaluation. Role-bearing agents do not only provide content. They enter settings in which users expect them to act as tutors, advisors, coaches, companions, or assistants. In those settings, accuracy is necessary but insufficient. The agent must also understand the purpose of the exchange, calibrate its authority, manage tone, and repair breakdowns when they occur.

The StudyChat analysis shows why this distinction matters. A capable model placed in a tutoring role and instructed only to be helpful can still fail interactionally. It may provide accurate content while doing too much of the student's work. It may understand the student's request while misjudging what the tutor role permits. It may respond fluently while failing to notice that the student needs a different kind of help. These failures are not captured by standard content-oriented evaluations, because the problem is not always what the agent says. It is whether the agent has performed the assigned role appropriately.

The practical conclusion is straightforward: *role-bearing agents need interaction specifications*. Builders should state what role the agent is occupying, what authority the role grants, what boundaries it must observe, what recurring situations it should recognize, and how it should respond when helpfulness becomes ambiguous. Evaluators should then score the agent against that specification, using role-specific rubrics and audit procedures that surface failures of purpose, authority, tone, and repair.

The aim is not to make agents more human. It is to make them more accountable to the roles they are already being asked to perform. As AI agents move into institutional and interpersonal settings, the evaluation infrastructure around them must expand beyond content accuracy and safety compliance. It must also ask whether the agent can sustain the interactional conditions of the role itself. Interaction Readiness names that missing layer and offers a first framework for building and evaluating it.